\documentclass[journal]{IEEEtran}

\usepackage{amsmath, amssymb,bm}
\usepackage{bbm}
\usepackage{url}
\usepackage{graphicx}
\usepackage{cite}
\usepackage{esvect}
\usepackage{subcaption}
\usepackage{amsthm}

\usepackage{booktabs}

\allowdisplaybreaks[1]
\allowdisplaybreaks
\usepackage{comment}
\usepackage{pifont}
\usepackage{color}

\ifCLASSINFOpdf
\else
\fi

\begin{document}
\title{Towards Scalable Wireless Federated Learning: Challenges and Solutions
%\thanks{}
}

\author{Yong Zhou, Yuanming Shi, Haibo Zhou, Jingjing Wang, Liqun Fu, and Yang Yang \\

\thanks{Y. Zhou and Y. Shi (corresponding author) are with ShanghaiTech University. H. Zhou is with Nanjing University. J. Wang is with Beihang University. L. Fu is with Xiameng University. Y. Yang is with Hong Kong University of Science and Technology (Guangzhou). 
}
}

\maketitle

\thispagestyle{empty}

\IEEEpeerreviewmaketitle

\begin{abstract}
The explosive growth of smart devices (e.g., mobile phones, vehicles, drones) with sensing, communication, and computation capabilities gives rise to an unprecedented amount of data. 
The generated massive data together with the rapid advancement of machine learning (ML) techniques spark a variety of intelligent applications. 
To distill intelligence for supporting these applications, federated learning (FL) emerges as an effective distributed ML framework, given its potential to enable privacy-preserving model training at the network edge. 
In this article, we discuss the challenges and solutions of achieving scalable wireless FL from the perspectives of both network design and resource orchestration. 
For network design, we discuss how task-oriented model aggregation affects the performance of wireless FL, followed by proposing effective wireless techniques to enhance the communication scalability via reducing the model aggregation distortion and improving the device participation. 
For resource orchestration, we identify the limitations of the existing optimization-based algorithms and propose three task-oriented learning algorithms to enhance the algorithmic scalability via achieving computation-efficient resource allocation for wireless FL. 
We highlight several potential research issues that deserve further study. 
\end{abstract}

%\begin{IEEEkeywords}
%Federated learning, 6G, over-the-air computation, learning to optimize.  
%\end{IEEEkeywords}

\section{Introduction}

With the commercial deployment of 5G systems worldwide, researchers from both academia and industry have kicked off to study the vision, drivers, use cases, and system requirements of 6G. 
Among many others, ubiquitous intelligence is envisioned to be a key feature of 6G and support various intelligent applications (e.g., smart city, autonomous driving) across the network coverage area. 
For intelligence distillation via centralized learning, an unprecedented volume of data generated by massive devices should be transferred to a central server, which may lead to large communication burdens and severe privacy leakages. 
Thus, edge artificial intelligence (AI) emerges as a promising technique for enabling scalable and trustworthy  model training at the network edge \cite{9606720}.

\begin{figure}[t] \centering
\includegraphics[scale = 0.33]{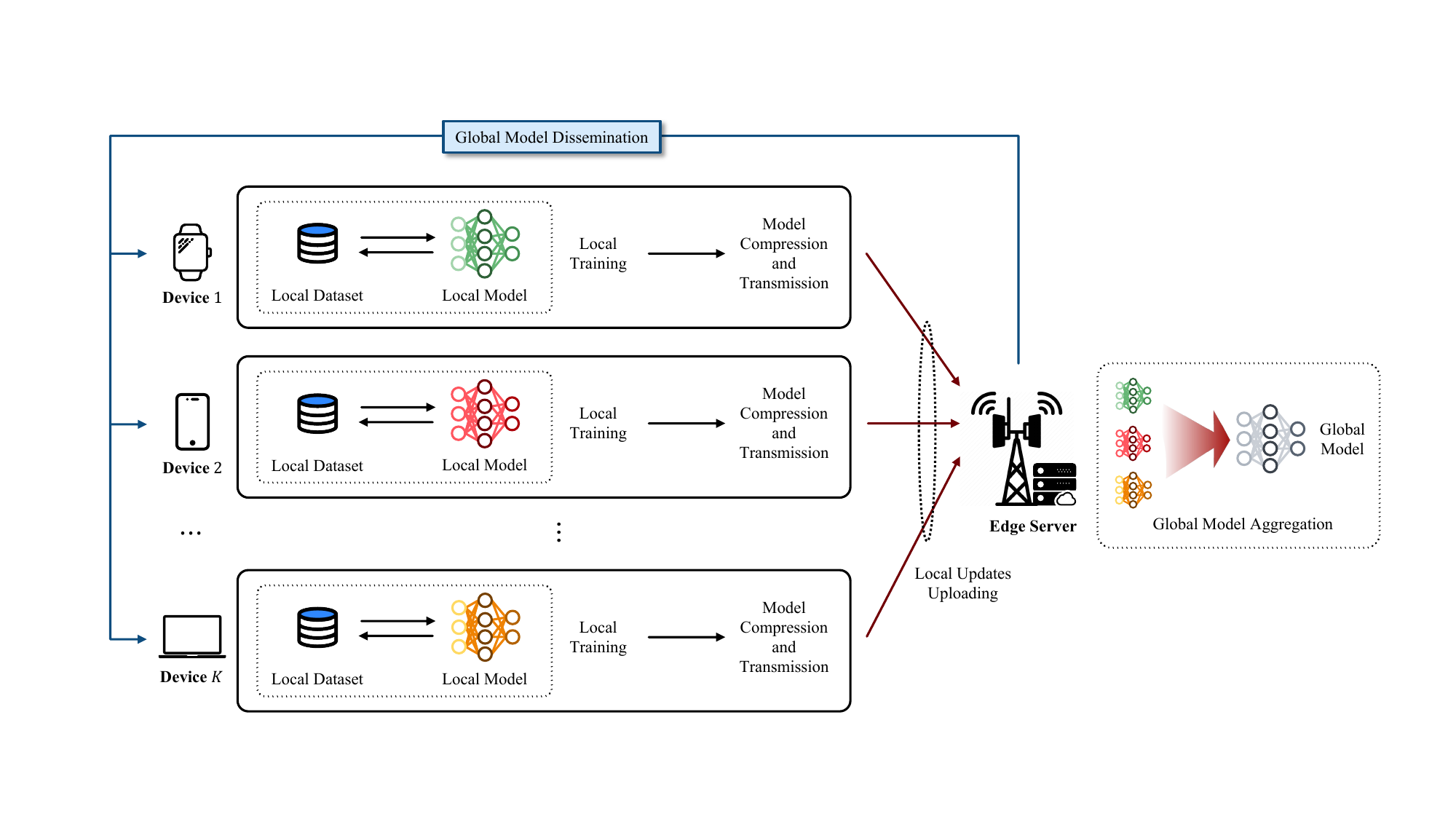} 
\caption{An illustration of the training process of FL.}
 \label{Fig:FL}
 \vspace{-2mm}
\end{figure}

Federated learning (FL) is recognized as an effective distributed machine learning (ML) framework, given its potential to enable privacy-preserving edge intelligence distillation while keeping the data locally \cite{WangTWC22}. 
According to the principle of FL, multiple devices collaboratively train a global ML model, where only the model parameters are communicated. 
Specifically, local training and model aggregation are two essential steps that are repeatedly performed, as shown in Fig. \ref{Fig:FL}. 
In the local training step, each device updates its local model according to the current global model and the local dataset. 
In the model aggregation step, an edge server aggregates the locally trained models to generate a new global model, which is then disseminated to the devices for the next round of training. 
By exploiting geographically distributed computing power and multi-modal data, FL pushes intelligence to the network edge \cite{9923620}. 

To facilitate wireless FL, the following three challenges need to be tackled. 
First, wireless FL relies on the frequent exchange of model parameters, which are usually high-dimensional. 
With limited radio resources, only partial devices can participate in each of the communication rounds, thereby limiting the amount of data to be exploited and degrading the learning performance. 
Second, because of wireless channel impairment, the model parameters transmitted over fading channels suffer from distortion that degrades the model aggregation accuracy. 
Third, time-varying wireless channels and dynamic network topology lead to the communication straggler issue, which reduces the learning efficiency. 
In general, more devices participating in FL facilitates the exploitation of more data, but aggravates the model distortion and communication straggler issues due to the occurrence of devices with poor channel conditions, which limits the communication scalability.
To break this dilemma, i.e., improving the communication scalability of wireless FL, it is essential to design task-oriented communication techniques \cite{TaskWCM, 9837474}.

Involving a large number of participating devices complicates the resource allocation due to the augmented dimension of optimization problems and the scarce radio resources, which demands efficient resource allocation.
Meanwhile, because of the unique features of wireless FL (e.g., multiple communication rounds), the conventional optimization-based resource allocation algorithms achieve a high spectrum efficiency at the cost of suffering from high computation complexity as well as poor algorithmic scalability and generalizability. 
Learning to optimize has the potential to tackle the limitations of optimization-based algorithms \cite{9562748}. 
Specifically, by exploiting the universal approximation capability of deep neural networks (DNN), learning-based algorithms establish a direct mapping between channel conditions and resource allocation decisions. 
However, data-driven learning algorithms have the limitations of relying on a high volume of training samples, converging slowly, and lacking of interpretability. 
This motivates the development of task-oriented learning algorithms that exploit the benefits of both optimization-based and data-driven learning algorithms to enable scalable resource allocation for wireless FL. 

In this article, we present a unified framework to enhance the scalability of wireless FL from two perspectives, i.e., network design and resource orchestration, by exploiting wireless for AI and AI for wireless, respectively.
	On one hand, we delineate the primary impediments to the performance  of wireless FL (e.g., inevitable model distortion, insufficient data exploitation), and develop effective wireless techniques to address these issues, thereby augmenting the scalability of task-oriented model aggregation.
	On the other hand, we put forth several task-oriented learning algorithms that exploit domain knowledge (e.g., mathematical model, network topology) and incorporate the goal (e.g., convergence optimality  and rate) into the algorithm design to enable computation-efficient resource allocation for wireless FL, thereby improving the algorithmic scalability.
	By unifying the scalable resource allocation techniques with scalable network design, our proposed framework has the potential to support both communication- and computation-efficient intelligent services.
We also highlight several open research issues that deserve further study.

\section{Network Design for Wireless FL}
\label{Wireless4AI}

In this section, we discuss how task-oriented model aggregation over wireless networks affects the performance of FL and propose several wireless techniques to improve  communication scalability. 

Model aggregation plays an essential role in distilling intelligence from multiple devices by aggregating their model parameters in each communication round. 
With appropriately designed model aggregation frequency and weights, many challenges of FL, e.g., communication bottleneck, system heterogeneity, and statistical heterogeneity, can be alleviated. 
However, due to wireless channel impairment and limited radio resources, implementing model aggregation over multiple-access fading channels suffers from inevitable distortion, including model aggregation errors due to deep fading and co-channel interference, and insufficient data exploitation due to limited device participation. 
These issues are critical as the accuracy of model aggregation and the amount of data exploited in each round determine the convergence optimality of FL. 
Below, we first elaborate how task-oriented communications can be applied for model aggregation and then present effective techniques that alleviate model aggregation distortion and improve data exploitation for enhancing communication scalability of wireless FL. 

\begin{figure*}[t] \centering
\includegraphics[scale = 0.54]{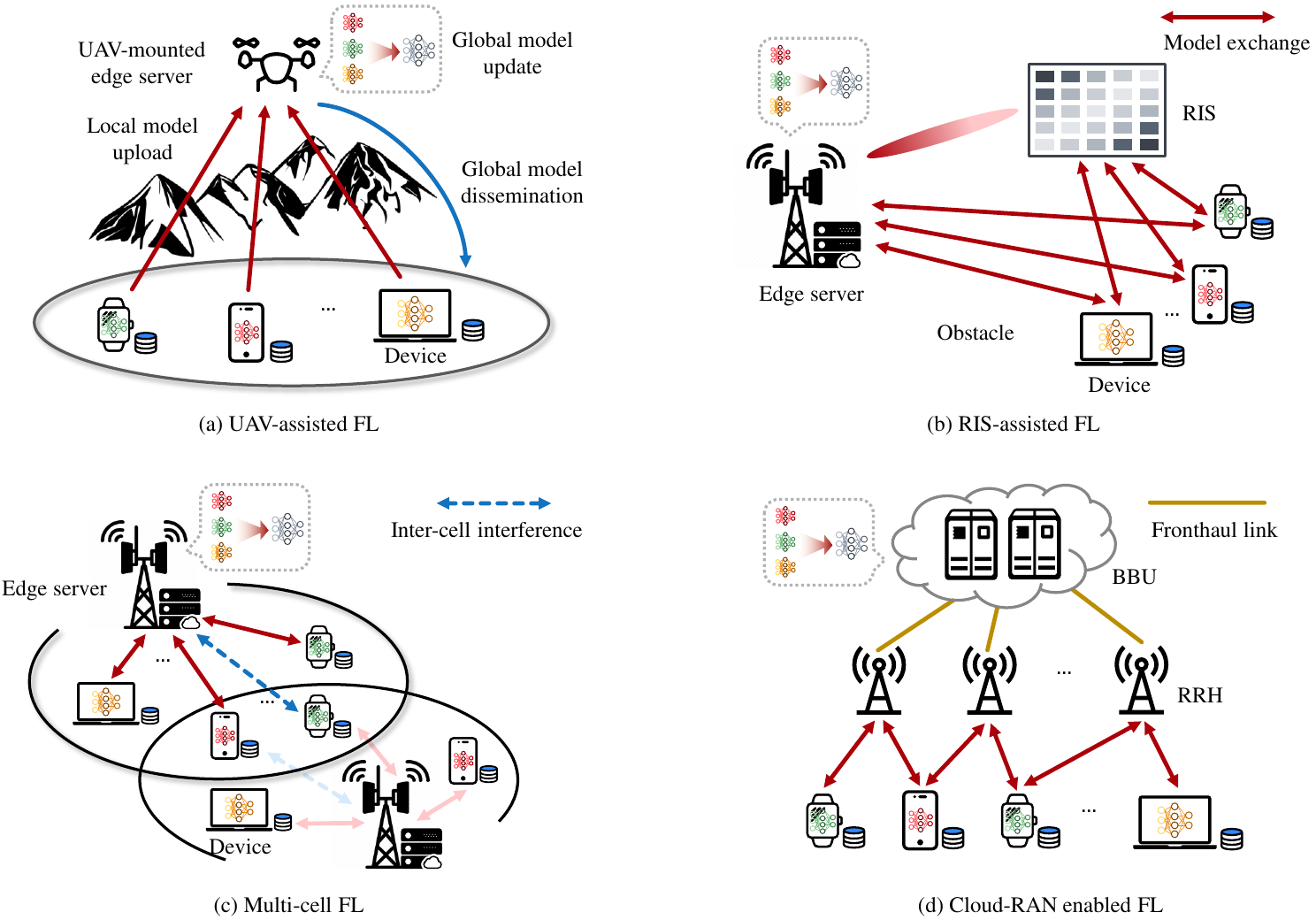}
\vspace{0mm}
\caption{Different network architectures for FL over wireless networks.}
 \label{Fig:1}
 \vspace{-3mm}
\end{figure*}

\subsection{Propagation Environment Reconfiguration} 

With conventional orthogonal multiple access schemes, the edge server performs model aggregation until the model updates from all participating devices are successfully received, which incurs high transmission delay. 
However, to perform the global model aggregation, the edge server only requires a specific function of local model updates, rather than each of the local model updates. 
Over-the-air computation (AirComp) as a task-oriented communication scheme receives much attention, given its capability of facilitating the concurrent transmission of multiple devices over the same channel and enabling the edge server to directly receive a specific aggregation of local model updates. 
Such a task-oriented model aggregation scheme ensures that the transmission delay does not scale with the number of participating devices. 
Note that time synchronization is essential for achieving the desired signal superposition via AirComp, which can be realized by advancing or retarding the signal transmission from different devices according to the timing advance technique.
Besides, a time-triggered synchronization strategy can be adopted to balance the computation and communication efficiency \cite{zhou2022timetriggered}.
However, for AirComp-assisted FL, the communication straggler enlarges the model aggregation error.
This is because all participating devices should adjust their transmit powers to align the signal magnitude at the edge server.
Hence, the devices with poor channel conditions, referred to as communication stragglers, lower the magnitude to be aligned at the edge server due to limited transmit power, thereby degrading the aggregation accuracy \cite{yang2020federated, 8970161}.
The dynamic network topology further aggravates the communication straggler issue and complicates the transceiver design. 
Fortunately, the communication straggler issue can be mitigated by reconfiguring the propagation environment. 
Unmanned aerial vehicle (UAV) and reconfigurable intelligent surface (RIS) are two promising techniques that can reconfigure the propagation environment by exploiting the mobility of UAVs and by introducing additional configurable propagation paths, respectively.

UAV can extend the network coverage area and provide broadband access in rural and underdeveloped areas by serving as mobile aerial edge server.
As shown in Fig. 2(a), with an UAV-mounted edge server, many devices can be prevented from being communication stragglers by enabling UAVs to proactively establish short-distance wireless links \cite{8918497}. 
However, as devices are usually geographically dispersed, the channel qualities of different uplink channels can be very different, which necessities the joint trajectory design and device scheduling for enhancing the performance of UAV-assisted FL.
By dynamically adjusting the UAV's location, each device has the opportunity to participate in FL, which enhances the learning performance by enlarging the data exploitation.
In general, scheduling more devices increases the transmission duration of each communication round, but reduces the number of communication rounds. 
Balancing such a tradeoff is essential for unleashing the full potential of UAVs to improve the scalability of task-oriented model aggregation.

RIS is capable of enlarging the received signal power by dynamically configuring its phase-shifts based on instantaneous channel state information (CSI).
With this salient feature, RIS can be leveraged to mitigate the communication straggler issue by reducing the channel heterogeneity among different devices, as shown in Fig. 2(b).
The tradeoff between the model aggregation error and number of participating devices is critical. 
Having more devices participating in model training increases the channel heterogeneity and leads to a greater model aggregation error. 
RIS can balance such a tradeoff, which, however, is challenging as the explicit relationship between these two aspects should be characterized, necessitating the task-oriented design. 
To tackle this challenge, the convergence analysis of RIS-assisted FL should be conducted and further exploited to formulate an optimization problem that maximizes the learning performance (e.g., minimizing the training loss, maximizing the test accuracy) of wireless FL.
Effectively solving such a problem helps in exploiting the benefits of RIS for enhancing the communication scalability \cite{WangTWC22}.

\subsection{Cooperative Interference Management}
The deep integration of wireless and AI technologies promotes the thriving of edge intelligence, making it normal to support multiple intelligent services over the same wireless network. 
To allow geographically dispersed devices receiving diverse intelligent services, a typical strategy is to deploy multiple edge servers to execute different FL tasks in a multi-cell network, as shown in Fig. 2(c), where all devices and edge servers share the same radio channel for uplink and downlink model exchanges.
However, the resultant inter-cell interference inevitably leads to model distortion during both the uplink model aggregation and downlink model dissemination processes as well as learning performance tradeoff among different FL tasks.
Under this circumstance, the inter-cell interference not only reduces the model aggregation accuracy but also limits the number of participating devices, and hence is a performance-limiting factor of scalable FL. 

Interference management should be able to balance the accuracy of model transmissions in different cells by not only aligning the intra-cell signals but also limiting the inter-cell interference.
However, interference management for conventional data transmissions (e.g., inter-cell interference coordination) cannot be directly applied to multi-cell FL due to the following challenges.
First, the metric for characterizing the performance of wireless FL is task-oriented and may not be explicitly expressed. 
Hence, the impact of inter-cell interference on the learning performance is difficult to be characterized. 
Second, the performance of wireless FL should be characterized from a long-term perspective, i.e., considering the impact of downlink/uplink model distortions accumulated over all communication rounds on the ultimate learning performance. 
Third, the achievable learning performance among different FL tasks is coupled. 
Hence, it is necessary to jointly design effective learning and communication schemes that account for task-oriented performance metrics and inter-cell interference.

\begin{figure}[t] \centering
\includegraphics[scale = 0.55]{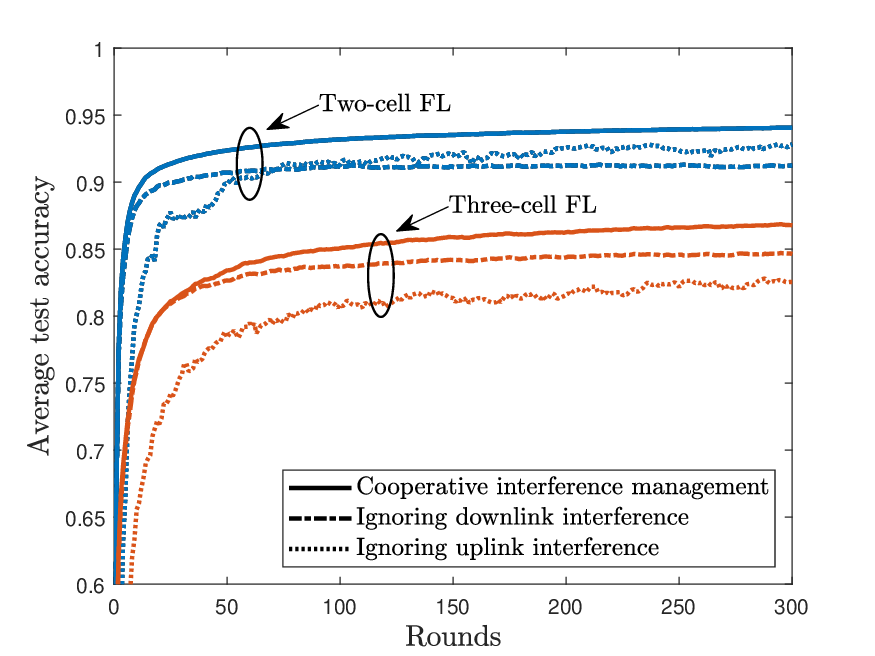}
%\vspace{-4mm}
\caption{Average test accuracy versus number of training rounds under cooperative interference management and non-cooperative schemes in both two-cell and three-cell FL systems. The cooperative design refers to the joint transceiver design among different cells for interference management and model distortion minimization, while the non-cooperative design refers to the separate design that ignores downlink/uplink inter-cell interference among different cells.  
} 
 \label{Fig:4}
 \vspace{-2mm}
\end{figure}

To facilitate interference management for multiple FL tasks, it is necessary to develop a cooperative optimization framework to balance the model aggregation accuracy achieved by different FL tasks, thereby enhancing the overall learning performance. 
Specifically, the downlink/uplink transmission distortions lead to a non-vanishing optimality gap, which detrimentally affects the convergence performance of all FL tasks \cite{WangJSAC22}. 
Intuitively, increasing devices' transmit power for one FL task is beneficial for reducing its model distortion, but enlarges the inter-cell interference to other FL tasks and in turn degrades their learning performance. 
Hence, the Pareto boundary in terms of the training loss due to model distortion can be adopted to characterize the performance tradeoff among FL tasks in different cells, thereby facilitating the cooperative design. 
As shown in Fig. \ref{Fig:4}, enabling cooperation among different cells can mitigate the detrimental impact of inter-cell interference, which in turn enhances the average test accuracy of multiple tasks in both two-cell and three-cell FL systems.
This demonstrates the effectiveness of cooperative interference management for supporting scalable wireless FL. 

\subsection{Cooperative Model Aggregation}

Although mitigating the communication straggler enhances the learning performance, the aforementioned techniques may not be applicable in large-scale wireless networks with massive dispersed devices that aim to learn a shared ML model. 
It is desirable to extend the network coverage area by deploying multiple edge servers. 
This not only shortens the communication distances to enhance the model aggregation accuracy, but also enables more devices to participate in FL and enlarges the amount of data to be exploited. 
However, it is challenging to enable cooperation among different edge servers while guaranteeing the communication efficiency.

By performing centralized baseband processing at the baseband unit (BBU) pool and deploying multiple remote radio heads (RRHs) as edge servers, cloud radio access network (Cloud-RAN) can be adopted as a multi-tier computing network to support cooperative model aggregation from massive devices. 
FL over Cloud-RAN consists of two phases to achieve cooperative model aggregation, i.e., the transmission of local updates from devices to RRHs via AirComp, and the transmission of aggregated signals from RRHs to the BBU via fronthaul links, as shown in Fig. 2(d). 
This procedure not only enlarges the coverage area using multiple RRHs, but also reduces the deployment cost by moving the baseband signal processing to the BBU.
However, both the model aggregation error due to AirComp and the quantization error due to limited fronthaul capacity degrade the model aggregation accuracy. 
Convergence analysis can be conducted to show that both the transmission distortion and quantization noise in the uplink and downlink prevent the FL algorithm from converging to the optimal solution. 
This necessities the investigation of how the model aggregation error degrades the learning performance as well as the joint optimization of communication and 
learning performance. 
Taking into account the limited transmit power and fronthaul capacity constraints, an efficient task-oriented resource allocation algorithm should be developed to minimize the optimality gap that reflects the detrimental impact of model distortion. 
With an appropriate design, Cloud-RAN enabled FL can achieve cooperative model aggregation to enhance the communication scalability by improving the data exploitation and reducing the model distortion.

\section{Resource Orchestration for Wireless FL} \label{AI4W}
To support wireless FL, it is essential to effectively allocate scarce communication and computation resources. 
This section focuses on computation-efficient resource orchestration for wireless FL and elaborates the advantage and design of task-oriented learning algorithms. 

\subsection{Knowledge-Guided Learning for Transceiver Design} 
\label{3A}
As FL is a long-term process, its convergence performance is determined by the accumulated model aggregation error. 
Hence, to enhance the convergence performance of AirComp-assisted FL, it is essential to minimize the time-average model distortion, while considering the dynamic channel conditions and the average transmit power constraints of devices. 
To this end, the devices' transmit powers and the receive beamforming vector (or the receive normalizing factor) of the multi-antenna (or single-antenna) edge server, i.e., AirComp transceiver design, should be jointly optimized, leading to a non-convex resource allocation problem. 

This long-term model distortion minimization problem can be decomposed into multiple per-round transceiver design sub-problems, which can be solved by developing an alternating optimization algorithm. 
However, such an optimization-based algorithm suffers from two limitations. 
First, it usually takes hundreds of iterations to converge to a stationary solution and has to solve a series of convex sub-problems in each iteration, which may incur a prohibitively high computation complexity. %, as FL typically involves a large number of edge devices. 
In addition, with time-varying channel conditions, this optimization-based algorithm should be executed in each channel coherence time, which further aggravates the computation complexity and reduces the training efficiency of FL. 
Second, the optimization-based algorithm relies on the CSI of all rounds to minimize the time-average model distortion. 
However, it is impossible to obtain the non-causal CSI (e.g., CSI of future communication rounds) in dynamic wireless networks. 

With the advancement of ML techniques, learning-based resource allocation algorithms can be developed to tackle the aforementioned limitations of optimization-based algorithms. 
Specifically, DNN can solve resource allocation problems by establishing a direct mapping between the instantaneous CSI and the AirComp transceiver design to support wireless FL. 
In general, learning-based resource allocation algorithms can be categorized into data-driven and model-driven learning algorithms. 
Although simple and widely adopted, data-driven learning algorithms take DNN as a black box and suffer from the limitations of relying on a high volume of training samples, converging slowly, and lacking of interpretability. 
As FL typically involves a large amount of devices and the corresponding resource allocation problem is high-dimensional, it is necessary to design task-oriented learning algorithms, where the neural networks are designed based on the domain knowledge to support both communication- and computation-efficient resource allocation for wireless FL, while achieving performance guarantee.

The mathematical expression of optimization variables can be exploited to develop model-driven learning algorithms for effectively solving resource allocation problems.
To minimize the time-average model distortion for AirComp-assisted FL, the optimal transmit power of each device can be explicitly expressed in term of the instantaneous CSI and dual variables by applying the Lagrangian-duality method \cite{zou2022knowledge}. 
The structure of the analytical expression of the transmit power can be exploited as domain knowledge to design a structure mapping layer when designing the neural networks. 
Hence, the task-oriented learning algorithm can be developed by setting the regularized time-average model distortion as the loss function for unsupervised learning with the DNN containing a knowledge-guided mapping layer.
Such a knowledge-guided design is capable of reducing the searching space of transmit powers and hence reducing the computation complexity, while achieving comparable performance as optimization-based algorithms. 
The knowledge-guided design only requires the instantaneous CSI to enable real-time resource allocation.

\subsection{Algorithm Unrolling for Sparse Gradient Recovery}
To enable collaborative training, high-dimensional model parameters need to be exchanged between devices and the edge server, which imposes a heavy communication burden. 
To alleviate this issue, various model compression techniques (e.g., sparsification, quantization) have been applied to achieve communication-efficient model transmission, without severely degrading the performance of FL. 
In particular, gradient sparsification exploits the fact that different entries of the local gradients have unequal contributions to the model update, where the elements with small magnitudes have little effects on the model convergence.
Meanwhile, most entries of gradients tend to be zero as the model training towards convergence, which further enlarges the gradient sparsity.
Therefore, gradient sparsification can be adopted to extract task-relevant information and reduce the parameter dimension for AirComp-assisted FL, where each device applies the Top-$K$ algorithm on the gradient and employs a random projection matrix before transmission. 
The recovery of the average gradient at the edge server is a compressive sensing problem, which can be tackled by adopting the approximate message passing (AMP) algorithms. 
However, the AMP algorithm performs in an iterative manner and may diverge when the projection matrix is ill-conditioned. 

\begin{figure}[t] \centering
\includegraphics[scale = 1.00]{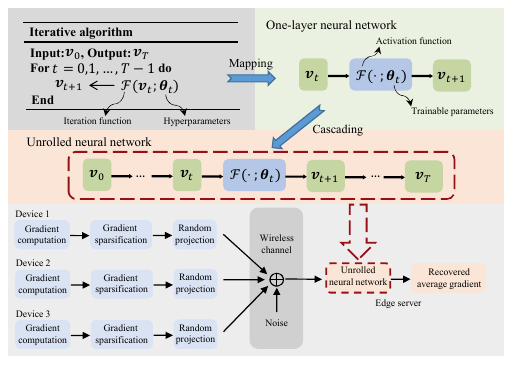}
\vspace{-1mm}
\caption{An algorithm unrolling architecture that first replaces each algorithm iteration with a neural network layer and then cascades these layers as an RNN, where $\mathcal{F}$ refers to the iteration (or activation) function, $\bm{\theta}_t$ refers to the hyperparameters (or trainable parameters) in the $t$-th iteration (or layer) of  iterative algorithms (or RNNs), $\bm{v}_t$ is the gradient to be recovered, and $T$ denotes the total number of iterations (or layers).}
 \label{Fig:5}
 \vspace{-1mm}
\end{figure}

\begin{figure*}[t] \centering
\includegraphics[scale = 0.55]{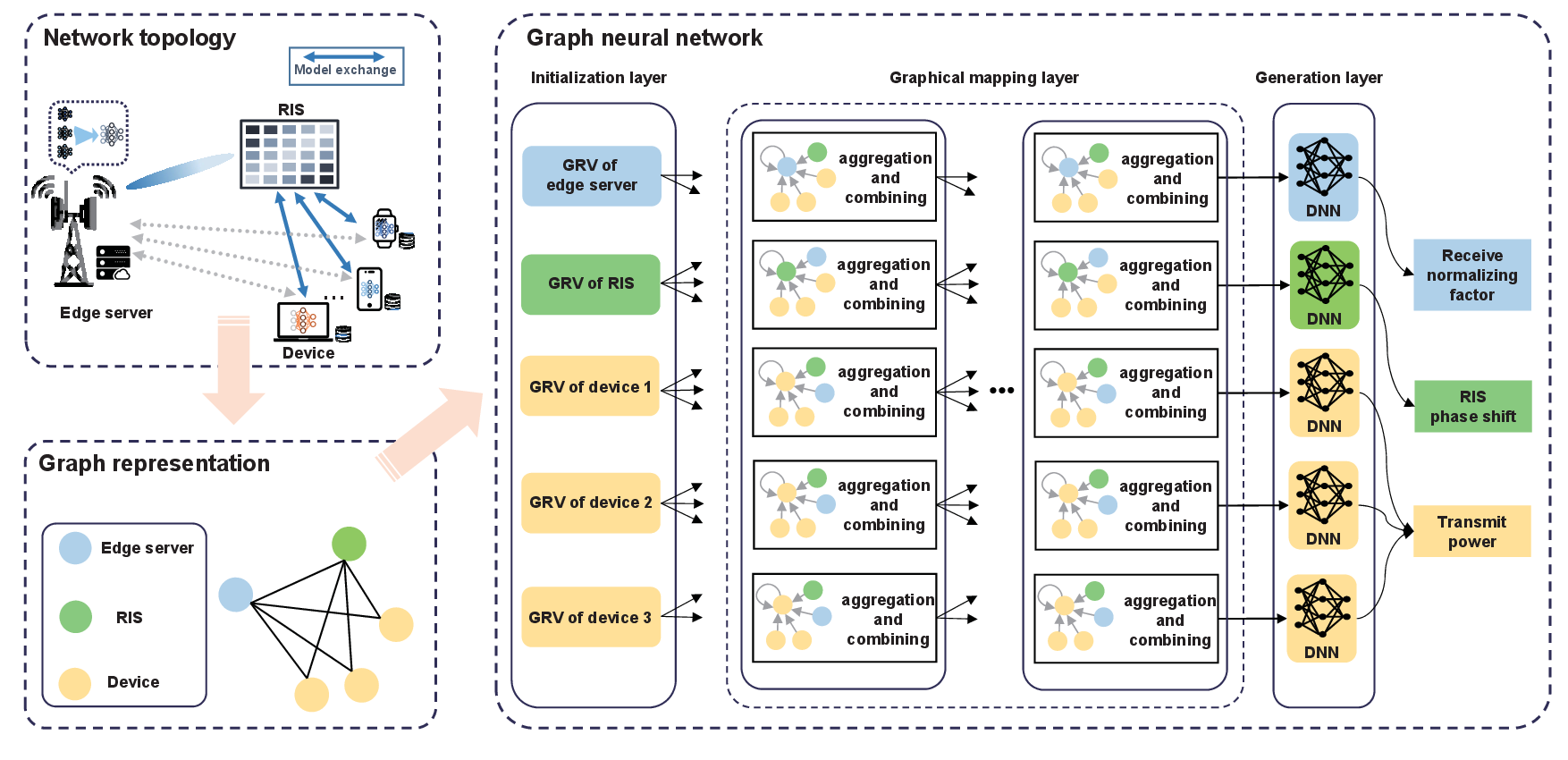}
\vspace{-0mm}
\caption{A GNN-based learning architecture for joint AirComp transceiver and RIS phase-shift design. 
The GNN-based learning architecture consists of the initialization layer that encodes the network topology and CSI into GRVs, the graphical mapping layer that updates the GRVs, and the generation layer that outputs the optimized receive normalizing factor, RIS phase-shifts, and transmit powers.}
 \label{Fig:3}
 \vspace{-1mm}
\end{figure*}

Algorithm unrolling is capable of transforming an iterative algorithm to a neural network \cite{monga2021algorithm}. 
One can replace each algorithm iteration with a neural network layer, and form a recurrent neural network (RNN) by cascading them layer-by-layer. 
This inspires us develop a task-oriented unrolling algorithm to unfold the AMP algorithm by incorporating the objective of recovering compressed model parameters into the RNN training, as shown in Fig. \ref{Fig:5}.
By exploiting the structure of iterative algorithms, the AMP-based unrolling algorithm can be applied for sparse signal recovery, thereby achieving both communication- and computation-efficient model transmission. 
The advantages of the algorithm unrolling for sparse gradient recovery are three-fold. 
First, the hyperparamaters need to be manually tuned for iterative algorithms, but can be directly optimized during neural network training for algorithm unrolling. 
Second, to solve high-dimensional resource allocation problems for wireless FL, the iterative algorithms are  computationally expensive, while the well-trained neural network via algorithm unrolling requires a low computation complexity in the inference stage. 
Third, algorithm unrolling, as a model-driven approach, inherits the structure of iterative algorithms and embeds the domain knowledge and the goal (e.g., convergence optimality and rate) into the neural network design. 
This not only reduces the computation complexity but also leads to excellent performance for wireless FL.

\subsection{GNN-based Learning for Joint Transceiver and RIS Design}
To leverage the full potential of RIS for mitigating the communication bottleneck and enhancing the model aggregation accuracy, the RIS phase-shifts and AirComp transceiver should be jointly optimized to minimize the time-average model distortion. 
Such a joint optimization problem is non-convex because of optimization variable coupling and unit modulus constraints introduced by the passive nature of RIS reflecting elements. 
This problem can be tackled by alternately optimizing the AirComp transceiver and the RIS phase-shifts, where the AirComp transceiver can be optimized by applying the Lagrangian-duality method. 
Additionally, the RIS phase-shifts can be optimized by transforming the original optimization problem to a rank-constrained semidefinite programming problem. 
The rank-one constraints can be tackled by applying the semidefinite relaxation technique, the difference-of-convex representation, or the Riemannian conjugate gradient technique. 
However, all the aforementioned methods generally incur a high computation complexity, especially when the devices and RIS phase-shifts are large in quantity.

With the capability of learning representation for network data and modeling complex data relationship, graph neural network (GNN) has shown its great potential in achieving scalable resource allocation \cite{9072356}. 
Wireless networks, consisting of many devices interconnected via wireless links, have intrinsic structures and can be well represented by graphs. 
This motivates the development of GNN-based learning algorithms that are capable of exploiting the network topology as a prior knowledge to efficiently solve the non-convex resource allocation problem for the aforementioned task, as shown in Fig. \ref{Fig:3}. 
With GNN, a graph representation vector (GRV) is defined for each node (e.g., edge server, RIS, and devices) to incorporate the local CSI. 
Subsequently, the GNN-based learning algorithm updates the GRVs via multiple graphical mapping layers that consist of multiple aggregation and combining modules to integrate information from other nodes, and generates the optimization variables via a generation layer.
The task-oriented GNN-based learning algorithm is capable of jointly optimizing AirComp transceiver and RIS phase-shifts to facilitate effective resource allocation for wireless FL, and outperforms the conventional fully connected neural networks in the following aspects. 
First, because of the representation capability of graphs and the unique combining and aggregation operations, GNN captures the interaction among the edge server, RIS, and devices, thereby enabling the joint optimization and improving the resource utilization for wireless FL. 
Second, because of the inherent permutation equivalence (PE) of GNN, the channel permutation leads to the same permutation of the transmit power of devices. 
With the permutation invariance (PI), the permutation of channels does not affect the optimization of RIS phase-shifts as well as the receive normalizing factor. 
These two properties significantly reduce the searching space of optimization variables and enhance the algorithmic scalability for resource orchestration. 
Third, each GNN layer consists of multiple aggregation and combining operations that share the same structures. 
This prominent feature ensures that the GNN-based learning algorithm is scalable to the number of devices, as only the number of modules needs to be adjusted and the neural networks do not need to be re-trained. 

\section{Open Research Issues}
In this section, we discuss several critical issues that deserve further study. 

\subsection{Integrating Sensing into FL} 

The learning performance of wireless FL heavily relies on the quantity and quality of data samples that are available at devices. 
Meanwhile, most of the related studies assume that the data samples are readily available. 
However, data sensing is a critical component in many practical FL applications, e.g., wireless sensing enabled human motion recognition, where each device needs to perform wireless sensing to obtain its local data samples \cite{9792281}. 
To incorporate wireless sensing into the framework of FL, many issues need to be considered. 
First, to obtain high-quality data, both effective waveform design and noisy sensing signal processing need to be studied. 
Second, as wireless sensing shares power and spectrum resources with task-oriented model updating and exchange, it is important to jointly optimize the resource allocation to enhance the overall learning performance. 
Third, real-time wireless sensing leads to the generation of streaming data, which should be considered for the design and analysis of wireless FL.

\subsection{Developing Trustworthy FL}

To facilitate the practical deployment of FL, it is critically important to develop effective schemes to ensure the privacy and security, thereby achieving trustworthy FL. 
Although FL mitigates the privacy concern to a certain extent by keeping the data locally, the exchange of model parameters may still leak privacy by applying model inversion attack and the global model may be intentionally manipulated by malicious devices. 
These issues motivate the development of effective privacy-preserving and secure model aggregation strategies to reduce the disclosure of local datasets and to defend the Byzantine attack, respectively. 
In this regard, many issues deserve further study. 
First, differential privacy introduces the random perturbations to the disclosed model parameters to enhance the privacy level, at the cost of enlarging the model distortion. Hence, it is necessary to develop effective task-oriented strategies to balance the tradeoff between the learning and privacy performance. 
Second, to defend the Byzantine attack, the edge server needs to perform non-linear model aggregation (e.g., geometric median), where AirComp cannot be directly applied. 
Developing a task-oriented communication scheme to achieve Byzantine-resilient model aggregation is necessary. 
Third, as differential privacy and Byzantine-resilience are two critical aspects of FL trustworthiness, it is essential to develop effective strategies that are not only robust against the Byzantine attack, but also capable of protecting the privacy of each device.

\subsection{Exploiting Diversified Datasets for Learning to Optimize}

To enable ubiquitous intelligence, future wireless networks shall support multiple intelligent services (e.g., FL tasks) over geographically dispersed cells. 
As the number of devices, radio propagation environment, and network topology of different cells are generally heterogeneous, each cell has to train a separate ML model to perform resource allocation based on the local dataset that is usually limited in size and only related to the local environment. 
Hence, the convergence rate, test accuracy, and robustness of each learned model for resource allocation may be limited. 
To mitigate these limitations, an effective method is to exploit diversified datasets, i.e., multiple cells collaborate to provide diversified datasets that can be used to train a high-quality ML model for resource allocation. 
As different cells may belong to different service providers, the direct sharing of data is prohibited. 
Under this circumstance, FL among different cells can be enabled, thereby enhancing the learning performance. 
For example, federated graph learning is a promising framework for the joint training of multiple GNNs to achieve effective resource allocation, while protecting the data privacy and enhancing the model quality. 
To implement federated graph learning, it is necessary to develop an effective method to mitigate the inherent heterogeneity of data distributions and graph structures in different cells.

\section{Conclusions}
FL over wireless networks, as a multidisciplinary research area that involves wireless communications, ML, and operation research, empowers many intelligent applications and gives rise to many new research opportunities. 
With an objective to enhance communication scalability of wireless FL, we advocated several wireless techniques to enhance task-oriented model aggregation and improve the device participation. 
Moreover, by exploiting the domain knowledge, we further proposed three task-oriented learning algorithms to enable computation-efficient resource allocation for wireless FL. 
Achieving effective network design and resource orchestration is an essential step towards scalable wireless FL.

\ifCLASSOPTIONcaptionsoff
  \newpage
\fi

\bibliographystyle{IEEEtran}
\bibliography{ref}

\vspace{3mm}
%\bibliography{reference}
%\bibliographystyle{ieeetr}

\end{document}